%% file: main.tex
\documentclass[sigconf]{acmart}
\AtBeginDocument{%
  }

\copyrightyear{2026}
\acmYear{2026}
\setcopyright{cc}
\setcctype{by}
\acmConference[CHIWORK Adjunct '26]{Adjunct Proceedings of the 5th Annual Symposium on Human-Computer Interaction for Work}{June 22--25, 2026}{Linz, Austria}
\acmBooktitle{Adjunct Proceedings of the 5th Annual Symposium on Human-Computer Interaction for Work (CHIWORK Adjunct '26), June 22--25, 2026, Linz, Austria}
\acmDOI{10.1145/3805029.3818299}
\acmISBN{979-8-4007-2598-2/2026/06}

\begin{document}

\title{Will AI Agents Free Us From Meaningless
Work? A Human-Centered Analysis}

\author{Davide Ghia}
\email{davide.ghia@polito.it}
\affiliation{
    \institution{Politecnico di Torino}
    \country{Italy}
}

\author{Jaspreet Ranjit}
\email{jranjit@usc.edu}
\affiliation{%
  \institution{University of Southern California}
  \state{California}
  \country{USA}
}

\author{Tania Cerquitelli}
\email{tania.cerquitelli@polito.it}
\affiliation{%
  \institution{Politecnico di Torino}
  \country{Italy}
  }

\author{Daniele Quercia}
\email{daniele.quercia@nokia-bell-labs.com}
\affiliation{%
  \institution{Nokia Bell Labs}
  \country{United Kingdom}
}

\renewcommand{\shortauthors}{Ghia et al.}

\begin{abstract}
Some claim that AI agents will free workers from the boring parts of their jobs, yet little is known about how workers themselves identify which tasks should be automated. Prior research focuses on occupations, overlooking that workers may experience varying levels of meaning across tasks within the same role.
We address this gap with a task-level analysis grounded in Graeber’s theory of bullshit jobs. Using ratings from 202 workers on 171 workplace tasks, we (1) validate a five-item scale of perceived bullshitness, (2) show that perceived bullshitness strongly predicts desire for AI delegation, and (3) find that such tasks are also seen as requiring less human oversight.
Together, these findings suggest that tasks perceived as bullshit are natural candidates for AI delegation, aligning worker preferences with perceived feasibility.
\end{abstract}




\maketitle


\input{1_introduction}
\input{2_related_work}
\input{3_dataset}
\input{4_analyses}
\input{5_discussion}
\input{6_conclusion}

\begin{acks}
    This study was partially supported by the ``WEBFARE'' (Nr. FISA2022-00908), funded by the Italian Ministry of University and Research under the FISA 2022 programme (D.D. No. 1405 of 13/09/2022, Fondo Italiano per le Scienze Applicate – FISA). This manuscript reflects only the authors’ views and opinions; the Ministry cannot be held responsible for them.
    
    \noindent\textbf{AI Disclosure statement.} ChatGPT was used for grammar and language refinement.
\end{acks}


\bibliographystyle{ACM-Reference-Format}
\bibliography{main}

\appendix

\end{document}

%% file: 1_introduction.tex
\section{Introduction}
\label{sec:introduction}

As AI systems become increasingly capable of acting autonomously, organizations face growing pressure to adopt them for productivity and cost reduction. These benefits dominate most discussions of AI in the workplace, while the effects on workers’ agency, satisfaction, and well-being receive far less attention \cite{VanderWeele2017}. As Graeber argues, many workers perform jobs they may experience as meaningless or unnecessarily bureaucratic, referred to as \emph{bullshit jobs}, and such perceptions can influence whether they wish to delegate these tasks to agentic AI systems or not \cite{Graeber2013,Graeber2018}.

A significant limitation of current research concerns the level of analysis. The relation between meaning and AI adoption is typically studied at the job or occupation level, although workers perform a diverse set of tasks within the same role \cite{Bailey2019,Shao2025}. Consequently, perceptions of meaningfulness can vary significantly within the same occupation.
For example, in our data, the same nurse rated the task ``Order, interpret, and evaluate diagnostic tests to identify and assess patient's condition'' as highly purposeful, while rating ``Record patients' medical information and vital signs'' as less meaningful.

If agentic AI systems are to be deployed responsibly, we must understand how workers experience tasks, and how these experiences relate to automation preferences and delegation feasibility. However, previous work lacks a validated way to measure perceived meaning at the task level, making it impossible to examine how it relates to AI acceptance or preferred AI agent behaviors.

To address this gap, we formulate three research questions: 
\begin{itemize}
    \item \textbf{RQ1}: What is perceived bullshitness at the task level, and how can we measure it?
    \item \textbf{RQ2}: How does perceived task bullshitness shape workers’ preferences for AI delegation and agentic behavior?
    \item \textbf{RQ3}: Are bullshit tasks also feasible targets for AI automation?
\end{itemize}

To address these questions, we make three contributions: first, we introduce and validate a task-level measure of perceived bullshitness, operationalizing workers’ subjective experience of tasks as pointless, unnecessary, or disconnected from meaningful outcomes; second, we show that perceived bullshitness strongly predicts workers’ preferences for AI delegation, providing empirical evidence that perceived meaning shapes what workers want to automate; third, we demonstrate that these same tasks are also seen as requiring less human oversight.

These patterns position perceived task meaning as a practical, human-centered measure to guide AI delegation decisions, complementing existing approaches that focus primarily on technical capability or task characteristics.

%% file: 2_related_work.tex
\section{Related Work}
\label{sec:related_work}

In recent years, meaningful work has become a central topic of debate, particularly with the spread of AI in the workplace \cite{Bailey2019,Microsoft2025}. Foundational work on human needs \cite{Maslow1943}, human flourishing \cite{VanderWeele2017}, and “bullshit jobs” \cite{Graeber2013,Graeber2018} highlights how lack of meaning negatively affects well-being, while recent studies link AI adoption to job anxiety through mechanisms such as Fear of Missing Out (FoMO) \cite{MendezSuarez2026}. However, these perspectives remain largely conceptual and do not provide empirical measures of meaning at the level of individual tasks.

In parallel, research on AI automation has focused on estimating exposure to technological change. Prior work uses O*NET task descriptions to assess how LLMs may reduce task completion time, typically reporting results at the occupation level \cite{Eloundou2023}. Similarly, the AI Impact Index maps AI innovations to O*NET tasks to estimate their impact across occupations \cite{Septiandri2024}. While informative, these approaches rely on externally defined task representations and overlook workers’ subjective experiences.

Recent work has shifted toward task-level analyses of AI impact. Autor et al. combine longitudinal labor-market data with patent-based measures to distinguish between automation- and augmentation-oriented innovations \cite{Autor2022}. Complementing this, recent worker-level evidence shows how AI exposure is associated with changes in tasks, skills, and wages \cite{Engberg2025}. Shao et al. introduce the Human Agency Scale (HAS) to measure preferred levels of human involvement in AI-assisted work \cite{Shao2025}, while another study investigates how workers make AI delegation decisions in practice \cite{Westphal2025}. More recently, Ranjit et al. \cite{Ranjit2026} examine how workers evaluate the meaningfulness of tasks exposed to AI, showing that highly exposed tasks are often associated with creativity, agency, and positive affect. Yet, these approaches do not capture how workers evaluate the desirability of automating specific tasks.

This leaves a gap between theories of meaningful work and task-level studies of AI impact. We address this gap by integrating theories of meaningful work with task-level measures of human agency and behavioral preferences for agentic AI systems. By capturing workers’ perceptions of “bullshitness” alongside their automation preferences, we provide a human-centered account of where and how AI can be deployed responsibly.

%% file: 3_dataset.tex
\section{Dataset}
\label{sec:dataset}
In this section, we show how we collected workers' ratings, describing how workplace tasks and participants were selected (Section \ref{subsec:task_selection}), and defining the survey items in detail (Section \ref{subsec:survey_composition}).

\subsection{Task selection and participants recruitment}
\label{subsec:task_selection}

We constructed a dataset capturing workers' task-level perceptions of bullshitness, automation preferences, preferred AI traits, and required human agency.

\noindent \textbf{Task filtering.} Candidate tasks were drawn from the O*NET occupational database \cite{ONET2025}, focusing on complex, multistep tasks associated with specific occupations rather than isolated actions. We restricted our selection to tasks that are primarily computer-based and plausible targets for AI agent automation or augmentation, following prior work on AI exposure at the task level \cite{Shao2025}. After this filtering, our dataset contained 10{,}131 tasks spanning 512 occupations.

\noindent\textbf{Screening process.} To ensure data quality, we first conducted a preliminary screening procedure. Using Prolific, we recruited U.S.-based participants belonging to one of 21 professional work functions and retained only those who rated at least one O*NET task in their occupation as highly familiar. We further retained only tasks rated as highly familiar by at least three participants. This process yielded a final sample of 202 workers evaluating 171 tasks across 22 occupations and 12 occupational sectors.

\noindent\textbf{Survey deployment.} Following the screening process, each participant rated up to 10 tasks that they had previously reported being highly familiar, providing task-specific responses across multiple survey blocks (Section \ref{subsec:survey_composition}). Task order within the survey was randomized to minimize order effects. In total, this procedure yielded 620 task ratings, corresponding to an average of 3.07 task evaluations per worker.

Participants had a mean age of 42.6 years (SD = 13.1). The sample includes 61.9\% participants identifying as female, 34.0\% as male, and 4.03\% did not disclose their gender; 46.94\% were employed full-time, 16.29\% part-time, and 4.03\% unemployed. Employment status was unavailable for 15.22\% of participants. Standard attention checks and familiarity thresholds were applied to exclude low-quality responses.

\subsection{Survey composition}
\label{subsec:survey_composition}

Workers evaluated each task using a structured survey composed of multiple task-level item blocks. The survey comprised 19 items capturing perceived bullshitness, automation preferences, and preferred AI agent behaviors, which form the basis of the analyses reported in the following sections.

\noindent\textbf{Perceived bullshitness.} We measured perceived bullshitness using five items adapted from Graeber’s theory of bullshit jobs \cite{Graeber2018}, capturing the extent to which a task feels: pointless (``The task feels pointless''), unnecessary (``If I stopped doing this task, nothing important would change''), disconnected from organizational goals (``This task does not contribute to the goals of my organization''), embarrassing (``I would be embarrassed to explain this task to someone outside my field''), or performed only for appearances (``I perform this task only to satisfy bureaucracy or appearances''). Items were rated on a five-point Likert scale. 

\noindent\textbf{Automation preferences.} We assessed automation preferences with two items adapted from Shao et al., measuring automation desire and required human agency when AI assists with the task \cite{Shao2025}. Finally, participants evaluated preferred AI behaviors through 12 agreement-based items capturing psychological traits that AI agents may exhibit \cite{Dong2024}. Each item represents a pair of polarized traits, such as ``Practical vs Imaginative'', or ``Polite vs Straightforward''. 

%% file: 4_analyses.tex
\section{Analyses}
\label{sec:analyses}
In this section, we first identify a bullshitness dimension based on the corresponding survey items (Section \ref{subsec:rq1}), then examine workers' preferences for bullshit tasks (Section \ref{subsec:rq2}), and finally assess whether those tasks are feasible targets for AI automation (Section \ref{subsec:rq3}).

\subsection{RQ1: What is perceived bullshitness at the task level, and how can we measure it?}
\label{subsec:rq1}

We derived a task-level measure of perceived bullshitness from survey items adapted from Graeber’s theory, capturing aspects such as purpose, meaning, and embarrassment (Section \ref{subsec:survey_composition}).

Exploratory Factor Analysis (EFA) revealed a single-factor structure (explained variance = 59.7\%) with high internal consistency (Cronbach’s $\alpha$ = 0.877). All items loaded strongly ($>$0.6) on the same latent dimension, supporting a unidimensional construct and the computation of a single bullshitness score, computed as the first-factor score from the EFA.

To interpret what this score captures in practice, we performed a qualitative analysis of tasks at the extremes of the score distribution. 
Tasks with low bullshitness scores often involve decision-making and produce concrete impact. These tasks shape outcomes rather than merely executing procedures, requiring expertise, interpretation, and evaluative judgment (e.g., ``Plan and direct special events for fundraising''). Another recurring pattern concerns human-centered coordination: tasks that support or coordinate human activity, contribute to collective goals, and involve social interaction are also frequently associated with low bullshitness (e.g., ``Confer with other staff members to plan and schedule lessons promoting learning''). 

In contrast, tasks with high bullshitness scores tend to be characterized by routine and repetitiveness. Procedural service tasks are highly standardized and require little discretion, limiting opportunities for agency or meaningful engagement from the worker’s standpoint (e.g., ``Process loan applications''). Administrative maintenance tasks are also prominent among high-bullshitness work: these tasks primarily exist to sustain institutional or legal systems rather than to directly help clients or achieve substantive outcomes (e.g., ``Prepare, draft, and review legal documents''), which may contribute to a perceived detachment from tangible human impact even if useful.

Having established a task-level measure of perceived bullshitness, we next examine how this perception shapes workers’ willingness to delegate tasks to AI.


\begin{figure*}[t]
    \centering
    \includegraphics[width=0.9\textwidth]{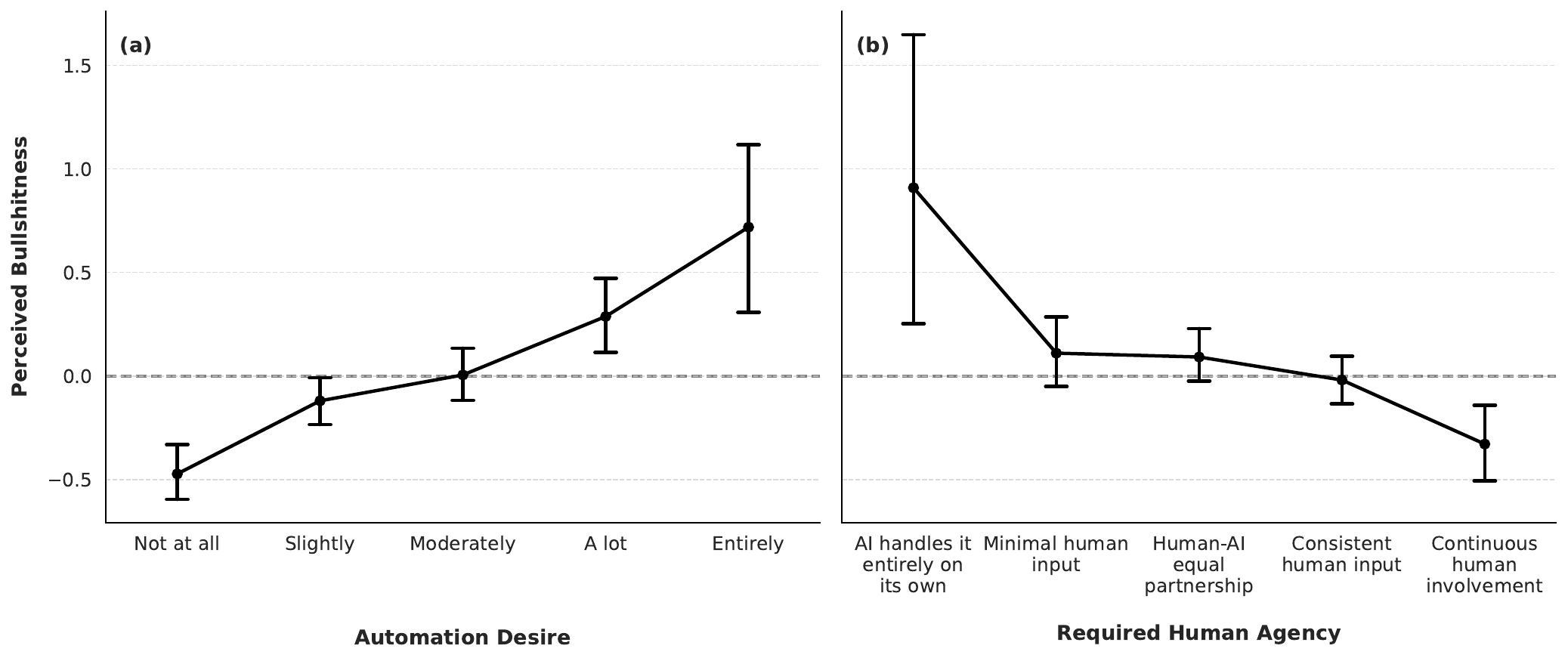}
    \caption{Perceived bullshitness in relation to (a) automation desire and (b) required human agency. Tasks perceived as more bullshit are associated with stronger preferences for automation and lower levels of required human agency, suggesting they are both desirable and feasible targets for AI automation.}
    \label{fig:bullshit_augm}
    \Description{
    Two scatterplots showing relationships between perceived task
    bullshitness and automation preferences. The left plot shows a positive relationship between bullshitness and automation desire: workers are increasingly willing to delegate tasks to AI as tasks are perceived as more meaningless. The right plot shows a negative relationship between bullshitness and required human agency: tasks perceived as more bullshit are associated with lower expected human involvement and greater AI autonomy. Both relationships follow approximately linear trends.
}
\end{figure*}

\begin{figure}
    \centering
    \includegraphics[width=0.98\columnwidth]{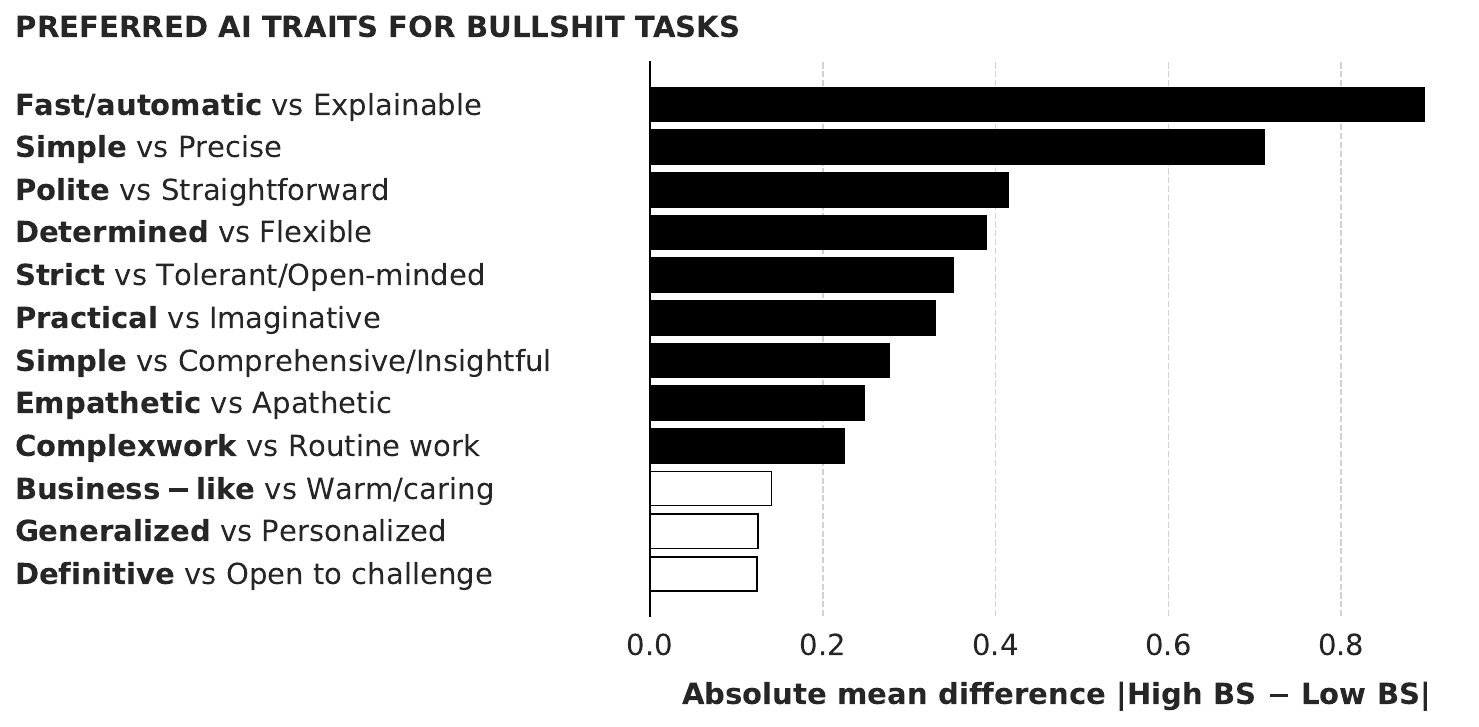}
    \caption{Preferred AI traits for tasks perceived as bullshit.
    Bars show the absolute difference in mean trait preference between tasks in the highest and lowest tertiles of perceived bullshitness. Each bar is labeled with the trait preferred for bullshit tasks (bold), while bar length indicates the strength of the preference shift. White bars denote traits for which the difference was not statistically significant. Overall, workers prefer AI agents that are fast, simple, practical for bullshit tasks, while still valuing politeness and empathy.}
    \label{fig:desired_AI_traits}
    \Description{
    Horizontal bar chart showing differences in preferred AI traits between tasks with high and low perceived bullshitness. The largest preference shifts for high bullshitness tasks favor fast and automatic, simple, practical, determined, and strict AI behavior. Preferences for politeness and empathy remain positive, while differences for explainability, flexibility, imagination, and
    precision are smaller or not statistically significant.
    }
\end{figure}

\subsection{RQ2: How does perceived task bullshitness shape workers’ preferences for AI delegation and agentic behavior?}
\label{subsec:rq2}

After defining perceived bullshitness at the task level, we examine workers’ willingness to delegate such tasks to AI. Following the approach of Shao et al. \cite{Shao2025}, we first analyzed responses to the item ``If an AI can do this task for you completely, how much do you want an AI to do it for you?'', with answers ranging from ``Not at all'' to ``Entirely''. 

A mixed-effects model with task and occupation as fixed effects showed that perceived bullshitness strongly predicts automation desire: increasing bullshitness by one standard deviation corresponds to an average increase of 0.39 points in automation desire on the five-point Likert scale ($\beta=0.39$, p < .001, 95\% CI [0.299, 0.480]). As shown in Figure~\ref{fig:bullshit_augm}a, automation desire increases monotonically with perceived bullshitness, following an almost linear trend. This suggests that workers are progressively more willing to delegate tasks to AI as those tasks are experienced as less meaningful or satisfying.

\noindent \textbf{Preferred AI traits.} Having established workers’ higher acceptance of automation for bullshit tasks, we next examine their preferences for specific agentic AI behaviors. Participants evaluated 12 bipolar items describing alternative AI traits. To support interpretation, we grouped tasks into tertiles based on perceived bullshitness and, for each trait, computed the difference in mean preference between tasks in the highest and lowest bullshitness tertiles (Figure~\ref{fig:desired_AI_traits}).

The results reveal a clear pattern: when automating bullshit tasks, workers favor AI agents that are fast, simple, practical, decisive, and rule-following. In contrast, traits associated with deliberation, exploration, or refinement, such as explainability, flexibility, imagination, or extreme precision, are less important in this context. Notably, politeness and empathy remain valued even for bullshit tasks, suggesting that workers may seek emotionally supportive or socially smooth AI behavior to mitigate the frustration associated with performing meaningless work.

Taken together, the trait-level results point to a consistent behavioral logic: when tasks are perceived as bullshit, workers care less about how the task is carried out and more about getting it out of the way. 


\subsection{RQ3: Are bullshit tasks also feasible targets for AI automation?}
\label{subsec:rq3}

After establishing that workers prefer to automate tasks they perceive as bullshit, we next examine whether such tasks are also perceived as suitable candidates for agentic AI automation.

Participants rated the level of human agency required to complete each task using the item “If AI were to assist in this task, how much of your collaboration would be needed to complete this task effectively?”, capturing the extent to which a task demands ongoing human judgment, initiative, or oversight \cite{Shao2025}. As shown in Figure~\ref{fig:bullshit_augm}b, perceived bullshitness is associated with systematically lower levels of required human agency: tasks rated as more bullshit are those for which workers believe AI can operate more autonomously, whereas tasks perceived as meaningful require more sustained human involvement.
This relationship is supported by a mixed-effects linear regression ($\beta = -0.216$, $p < .001$, 95\% CI [-0.303, -0.130]), with task and occupation included as fixed effects.

This alignment suggests that workers’ preference for automating bullshit tasks reflects not only subjective desirability, but also perceived feasibility, highlighting these tasks as promising targets for AI adoption.



%% file: 5_discussion.tex
\section{Discussion}
\label{sec:discussion}

Our findings suggest a clear implication for the design of agentic AI systems: tasks that workers perceive as more bullshit are particularly suitable targets for automation. Perceived bullshitness can serve as a task-level heuristic to dynamically configure agent autonomy and interaction style. In operational settings, it may also support monitoring and intervention: tasks expected to require minimal human involvement could flag system issues when frequent overrides occur, enabling targeted reconfiguration of agent behavior.

Ranjit et al. show that the most meaningful tasks involve high levels of creativity, novelty, agency, and a clear contribution to organizational goals \cite{Ranjit2026}. In line with these results, we find that tasks with the lowest bullshitness scores are those that involve decision-making, produce concrete impact, and contribute to collective goals. While their work characterizes meaningfulness through multiple positive dimensions, our results provide a complementary perspective by operationalizing its absence: tasks perceived as “bullshit” are precisely those workers prefer to delegate to AI, and that require minimal human agency.

We emphasize that perceived bullshitness is context-specific: workers' subjective perceptions may vary substantially across cultures, organizations, or historical periods. For this reason, we do not propose this concept as a universal truth but as an audit lens that complements existing automation criteria. Decisions about AI deployment often prioritize technical feasibility or organizational requirements; our findings suggest that workers' preferences regarding which tasks they wish to retain should also be considered. Furthermore, perceived bullshitness is not a synonym for a task's objective value. Tasks that workers experience as disconnected from meaningful outcomes may nevertheless serve important organizational functions, such as compliance or accountability. Our results therefore do not imply that such tasks should be eliminated. Rather, if AI systems can perform them reliably, automation may preserve these functions while allowing workers to devote more attention to tasks they find meaningful.

A complementary pattern concerns how workers expect AI to handle tasks perceived as meaningless. Preferences consistently favor efficiency-oriented behavior, prioritizing speed, simplicity, and minimal interaction. This suggests that, for highly bullshit tasks, agent design should emphasize constrained, efficient execution rather than explainability or creativity. More broadly, task-level perceptions of meaning can inform early design choices, including the use of foundation models, tools, or hybrid approaches.

\noindent \textbf{Limitations}. Our findings should be interpreted in light of several limitations. First, participants were recruited exclusively from the United States, which may limit generalizability. Second, we focus on task characteristics and do not model contextual factors such as organizational practices or compensation structures. Third, we assess perceived rather than technical feasibility of automation. Finally, our results reflect a snapshot in a rapidly evolving technological landscape, and preferences may shift as AI capabilities develop.

\noindent \textbf{Future work}. Future research could develop standardized frameworks to identify bullshit tasks across contexts and support responsible AI deployment decisions. Extending beyond bullshitness, additional dimensions of meaningful work, such as psychological well-being or social contribution, could be examined in relation to automation preferences. Finally, combining worker perceptions with expert evaluations could provide a more comprehensive view of both desired and feasible AI delegation.

%% file: 6_conclusion.tex
\section{Conclusion}
\label{sec:conclusion}

We introduced a task-level measure of perceived bullshitness and validated it as a unidimensional construct. Using this measure, we found that tasks perceived as bullshit are both highly desirable and suitable targets for agentic AI systems.

Our analyses further reveal that workers’ expectations for AI behavior depend strongly on task meaning: when automating meaningless tasks, workers favor agentic systems that are fast, simple, practical, and decisive, rather than explainable, flexible, or exploratory. 

By grounding automation decisions in workers’ lived experiences of meaning, agency, and collaboration, our work offers practical guidance for deploying agentic AI systems in ways that improve productivity without undermining the meaningful aspects of human work.